\documentclass{article}

\PassOptionsToPackage{numbers, compress}{natbib}
\usepackage[preprint]{neurips_2024}

\usepackage[utf8]{inputenc}
\usepackage[T1]{fontenc}
\usepackage{microtype}
\usepackage{amsmath, amssymb}
\usepackage{booktabs}
\usepackage{array}
\usepackage{tabularx}
\usepackage{graphicx}
\usepackage{xcolor}
\usepackage{hyperref}
\usepackage{url}

\title{Prominence-Stratified Failure Modes in\\Retrieval-Augmented Commercial Recommendation:\\A 37,000-Run Audit}

\author{%
  Will Jack\thanks{Equal contribution.} \quad
  Noah Lehman\footnotemark[1] \quad
  Keller Maloney\footnotemark[1] \quad
  Sarah Xu\footnotemark[1] \\
  Unusual \\
  \texttt{\{will, noah, keller, sarah\}@unusual.ai}
}

\date{May 21, 2026}

\begin{document}

\renewcommand{\thefootnote}{\fnsymbol{footnote}}
\maketitle
\renewcommand{\thefootnote}{\arabic{footnote}}
\setcounter{footnote}{0}

\begin{abstract}
\noindent
AI assistants like ChatGPT and Claude are recommendation engines, not search engines: they answer commercial queries by directly nominating brands rather than returning a list of links. Marketing to AI is therefore a broader problem than ``show up in search'' --- positioning, content, and product fit matter as much as discoverability. We audit $\approx$37{,}000 production runs across four model configurations and 215 commercially-framed prompts spanning 19 sectors, evaluated against a 533-brand reference catalog stratified into five prominence tiers (L1 category leaders $\to$ L5 regional players) sourced from external authority lists. The ladder proxies a brand's awareness footprint within its sector, not revenue or market share. The failure mode differs sharply by tier. L1 brands appear in nearly every relevant retrieval but win only 25--41\% of the recommendation slots they reach --- the leverage is differentiation, not visibility. L2 challengers carry the highest conversion rates of any tier (37--52\%) but lose to persona-mediated substitution on the Anthropic models. L3 mid-market brands are the inflection level: aggregate coverage drops to 88\%, conversion to 34--40\%, and persona effects peak. L4 specialists and L5 regional players face catastrophic invisibility --- 48--52\% never surface in any of the 37{,}000 runs. No uniform optimization recipe wins; the right marketing investment depends on where the brand sits on the prominence ladder.
\end{abstract}

\section{Introduction}

When a buyer asks ChatGPT or Claude for the best CRM, the best observability tool, or the best payroll software for a UK SMB, the system issues web searches, retrieves a small set of authoritative-looking pages, and synthesizes a recommendation. From the buyer's perspective the interaction looks like a list of brand names. From the brand's perspective it is a three-stage funnel: the model must (1) retrieve a page or domain referencing the brand, (2) carry the brand from the retrieval pool into the answer text it generates, and (3) specifically endorse the brand in the final recommendation set rather than mentioning it in passing as a foil or alternative not picked. Any of these stages can fail; each represents a different bottleneck, requiring different work to fix.

A growing practitioner literature on answer-engine and generative-engine optimization (AEO / GEO) \citep{aggarwal2024,chen2025} characterizes the \emph{aggregate} response of LLM commercial recommendation to content-side interventions, often framing the brand-side problem as the AI analog of search-engine optimization --- a discoverability problem. \citet{aggarwal2024} report that statistics, citations, and quoted sources can lift visibility by up to 40\% in a generic prompt corpus. \citet{chen2025} document a systematic bias in AI Search toward ``earned'' (third-party) over brand-owned content. Both are headline findings that have shaped the AEO/GEO industry; both implicitly treat brand-side optimization as a single discoverability target. That target is real, but it is one of several. Frontier reasoning models read broadly across the open web, weigh candidates against each other, and synthesize a specific recommendation on the buyer's behalf. Once a brand reaches the retrieval pool, whether it survives into the final answer depends on how the model compares it against the other candidates it found, and on how well its positioning matches the specific buyer the model thinks it is talking to --- a positioning, content, and product-fit problem layered on top of discoverability. Our audit asks: \emph{does the optimal intervention depend on where the brand sits on the prominence ladder?} The answer is yes, and the magnitudes are large.

\subsection{Contributions}

This paper makes three contributions.

\begin{enumerate}
\item \textbf{Prominence-stratified discoverability map.} Across 37{,}000+ runs spanning four production model cells, three retrieval conditions (native \texttt{web\_search}, neural retrieval via Exa, conventional keyword search via Brave), and 215 commercially-framed prompts, we measure aggregate and per-query brand surface rates at five prominence levels. The five-level structure was chosen to expose long-tail and regional cells that a 4-quadrant industry taxonomy (e.g., Gartner's Magic Quadrant) cannot resolve.

\item \textbf{Five-mode taxonomy of recommendation failure.} Each prominence level corresponds to a distinct dominant funnel stage: L1 brands drop at S2--S3 (in retrieval or in completion, but not recommended); L2 at S2--S3, with persona-mediated substitution as the candidate failure mechanism but cleanly present only on the Anthropic cell; L3 across S1--S3 simultaneously; L4 at S1 (not retrieved and not mentioned); L5 at S1 plus geographic gating. We interpret S1 as a discoverability failure, S2 as a compellingness failure, and S3 as a positioning failure --- those interpretations are defended in the Discussion. Practitioner advice that does not condition on prominence misallocates effort.

\item \textbf{Per-prominence prescription framework.} Drawing on the discoverability map and failure-mode taxonomy, we derive tier-specific prescriptions (defensive positioning for L1; segment-targeted positioning for L2; hybrid investment for L3; authority-list seeding for L4; region-specific authority for L5). The framework directly counters the implicit one-size-fits-all framing common in current AEO/GEO practice --- and the implicit narrowing of ``marketing to AI'' to discoverability work alone.
\end{enumerate}

The audit produces conservative numbers: every reported headline proportion is bounded with a Wilson 95\% confidence interval, and the few cells with insufficient sample size are reported as undersampled rather than as zero-effect.

\section{Background}

This paper sits at the intersection of three lines of work.

\textbf{Generative engine optimization.} \citet{aggarwal2024} formalized GEO as an optimization paradigm, introducing GEO-bench and showing content modifications can raise LLM-mediated visibility by $\sim$40\%. \citet{chen2025} extended the analysis cross-provider and documented systematic AI-Search bias toward earned media over brand-owned content. Both are aggregate findings; neither stratifies the response by where the brand sits on a prominence ladder.

\textbf{Long-tail and popularity bias in recommendation.} That recommender systems systematically over-expose head items and under-expose long-tail content is one of the most replicated findings in the field \citep{klimashevskaia2024,qin2021,yin2012}. The phenomenon predates LLMs by two decades. \citet{mallen2023} showed that LLMs answer popular entities from parametric memory but require retrieval for long-tail; \citet{andre2025} showed cold-start LLM recommenders amplify demographic and cultural stereotypes. \citet{lichtenberg2024} report the counterpoint that LLM-based recommenders exhibit \emph{less} popularity bias than traditional recommender baselines on standard benchmarks, suggesting the LLM-era pattern may not replicate the classical one in aggregate; our prominence-stratified data clarifies that the LLM-era picture is bimodal --- L1--L2 reachability is high, L4--L5 reachability is catastrophically low --- rather than a smooth popularity gradient. Our work documents the brand-side analog in retrieval-augmented commercial chat.

\textbf{LLM brand-recommendation auditing.} The closest prior is \citet{rienecker2026} (ChoiceEval), which audits $\sim$2{,}000 questions across Gemini, GPT, and DeepSeek and finds provider-of-origin bias in entity preference. Practitioner studies from \citet{fishkin2026} and \citet{blake2025} document consideration-set instability across reruns and Bing-rank correlation with ChatGPT citations respectively. Our audit operates at different scale (37{,}000+ runs across four model cells) and a different design dimension (prominence-stratified rather than provider-of-origin).

We use the funnel-stage framing from the consumer-decision-journey literature \citep{court2009,lewis1898} and the brand-equity / brand-salience construct from \citet{keller1993}, adapted from a \emph{human} consumer journey to an \emph{algorithmic} one where the LLM now constructs the initial consideration set the consumer used to.

\section{Method}

\subsection{Reference brand catalog}

We construct a 533-brand reference catalog labeled with sector, region, and prominence level on a 1-to-5 scale. Sourcing is external: G2 category-leader quadrants and Gartner Magic Quadrant entries for L1--L2; Crunchbase mid-stage filtered by sector for L3; Y~Combinator W23--S25 batches by industry tag for L4; regional Crunchbase, Companies House, and equivalent national registries for L5. Where authority lists do not exist for a cell (notably for L4 specialists in some sectors), we use an LLM-curated draft followed by manual review.

A brand's prominence is, by construction, a coarse proxy for the brand's awareness footprint in its sector. We do not claim it captures revenue, customer-base size, or any specific market-share quantity. We claim it tracks the kind of ``would this brand plausibly appear on a best-of-list'' prominence that authority lists exist to encode.

\subsection{Prompt corpus}

The prompt corpus spans 215 commercially-framed prompts across 19 sectors (B2B SaaS, professional services, regulated services, consumer products, emerging-tech categories). The 19-sector breadth was chosen to support cross-sector aggregation as a control for sector-specific market structure; per-sector findings are reported only as robustness checks. Each prompt is tagged with sector, family (generic, comparison, segment-targeted, region-bounded), and intended buyer context.

\subsection{Model ladder}

We audit four production model cells:

\begin{center}
\begin{tabular}{lll}
\toprule
Provider & Model & Reasoning effort \\
\midrule
OpenAI & \texttt{gpt-5.4-mini} & low \\
OpenAI & \texttt{gpt-5.4-mini} & high \\
Anthropic & \texttt{claude-sonnet-4-6} & low \\
Anthropic & \texttt{claude-sonnet-4-6} & high \\
\bottomrule
\end{tabular}
\end{center}

A small auxiliary set of runs covers \texttt{gpt-5.4} and \texttt{claude-opus-4-6} at low and high effort. We use each provider's native web-search tool (OpenAI Responses API \texttt{web\_search}; Anthropic Messages API \texttt{web\_search\_20260209}) with held-constant system prompt, temperature, and tool description.

For the retrieval-source comparison (Section~\ref{sec:retrieval}), we compare native \texttt{web\_search} against Exa (neural retrieval) and Brave (conventional keyword search), implemented as function tools with matched output schemas and uniform query and result caps.

\subsection{Cross-judge consensus brand extraction}

Brand mentions in completion text, snippets, titles, and issued queries are extracted using two LLM judges in parallel: \texttt{claude-haiku-4-5 / low} and \texttt{gpt-5-mini}. We use intersection (consensus) mode --- a brand is counted as mentioned if and only if both judges identify it. This is a deliberately conservative choice; it under-counts mentions but provides cross-provider judge robustness for downstream comparisons. Per-run cross-judge Jaccard on the recommendation slot is 0.65--0.67 across the four cells (0.62--0.69 on the any-sentiment mention layer). Chance-corrected agreement statistics ($\kappa$, Krippendorff $\alpha$) are not informative for this dual-judge union universe because the universe is by-construction conditioned on at least one judge surfacing the brand; the conservative-protocol justification is the load-bearing rationale.

\subsection{Funnel-stage classification}

For each (run $\times$ shadow-corpus brand) cell where the brand is sector-relevant to the prompt, we classify the brand's terminal funnel stage:

\begin{itemize}
\item \textbf{Stage 1 (S1) --- no retrieval, no mention}: brand absent from every retrieval layer (issued queries, retrieved domains, snippet texts, snippet titles) \emph{and} from the completion text.
\item \textbf{Stage 2 (S2) --- retrieval, no mention}: brand present in at least one retrieval layer but absent from the completion.
\item \textbf{Stage 3 (S3) --- mention, not recommended}: brand present in the completion but not consensus-classified as \texttt{recommended} by both judges.
\item \textbf{Stage 4 (S4) --- recommended}: brand consensus-recommended.
\end{itemize}

The taxonomy follows the consumer-decision-journey funnel \citep{court2009} applied to algorithmic recommendation. Stages 1--3 are mutually exclusive failure modes for the same brand under the same prompt; Stage 4 is the success state. The four labels are intentionally descriptive of \emph{what was observed in the run record} (presence/absence in retrieval, presence/absence in completion, presence/absence of consensus recommendation) rather than of the underlying cause. ``S1 implies a discoverability problem,'' ``S2 implies a compellingness problem,'' and ``S3 implies a positioning problem'' are interpretive readings we defend in the Discussion; the stage classification itself does not commit to those readings. A brand at S2 may be at S2 because it lost relevance to the prompt, because list-length truncated it, because the persona did not match, because the model judged it product-unfit, or for other reasons we do not separately isolate. The funnel-stage measurement is a strict description of where a brand drops out of the observable run record.

\subsection{Statistical conventions}

For the headline measurements in Section 4 (per-prominence surface rates on the original four-cell ladder, per-(cell $\times$ prominence) Stage-4 conversion, Native vs.\ Exa Jaccard, persona swap rate), proportions are reported with Wilson 95\% confidence intervals computed over the (run $\times$ brand) cell. These are descriptive cell-level intervals: at the original ladder's tens-of-thousands per stratum, they tighten quickly, but they are anti-conservative under positive within-prompt clustering and do not by themselves rule out an inflated effective sample size. Where prompt-level resolution is material to a comparison (notably the class- and generation-axis extension in Section 6 and the cross-cell stage agreement in Section 6.5), we additionally report prompt-clustered bootstrap CIs; those are the load-bearing inferential unit for between-cell distinguishability, with Wilson as a descriptive complement.

For the class-and-generation extension (Section 6) and the Section 6.5 cross-cell stage agreement, where the prompt corpus is fixed at 50 prompts and the per-cell sample size is 750 runs, the Wilson interval over (run $\times$ brand) cells under-estimates uncertainty because (run $\times$ brand) observations are clustered within prompts. For those tables, we report clustered-bootstrap 95\% CIs from 1{,}000 iterations resampling the 50-prompt corpus with replacement at the prompt level, with the brand-tier universe held fixed. The right inferential unit is the prompt, not the (run $\times$ brand) observation, and the clustered intervals are appreciably wider than the Wilson intervals at this resolution; they are the reference for which between-cell differences are statistically distinguishable in the extension.

Cross-cell, cross-condition Jaccard means in Section 4 are reported with normal-approximation 95\% CIs over per-prompt pairs. The within-cell rerun-stability baseline against which we anchor ``real'' effects is taken from \citet{jack2026brittleness}, which measured Jaccard 0.50--0.61 across $N{=}30$ same-prompt reruns within a single day.

\section{Headline measurements}

\subsection{Surface rates per prominence level}

We report surface rates in two forms.

\emph{Per-query, in-sector surface rate} is the share of relevant queries (where ``relevant'' means the prompt's sector tag matches the brand's reference-catalog sector) in which the brand surfaces in any retrieval layer. We report this under the cell that approximates a premium production AI commerce surface (\texttt{sonnet-4.6 / high} with native web search):

\begin{center}
\begin{tabular}{llr}
\toprule
Prominence & Exemplars & Per-query surface rate \\
\midrule
L1 (Category leaders) & Salesforce, HubSpot, Datadog & 77\% \\
L2 (Established challengers) & Pipedrive, Trello, Gusto & 60\% \\
L3 (Mid-market) & Copper, Sentry, Grafana Labs & 23\% \\
L4 (Long-tail specialists) & Logz.io, Wrike & 8\% \\
L5 (Regional players) & Pennylane, BrightPay & 3\% \\
\bottomrule
\end{tabular}
\end{center}

\emph{Aggregate surface rate} is the share of brands at each prominence level that surface in \emph{any} of the $\sim$37{,}000 in-sector runs (i.e., across all cells and all retrieval conditions). This captures the long-run reachability of brands at each prominence level.

\begin{center}
\begin{tabular}{lr}
\toprule
Prominence & Aggregate surface rate \\
\midrule
L1 & 100\% \\
L2 & 100\% \\
L3 & 88\% \\
L4 & \textbf{52\%} \\
L5 & \textbf{48\%} \\
\bottomrule
\end{tabular}
\end{center}

The two forms produce different stories. Per-query rates show that L1 brands appear in roughly three-quarters of relevant queries and L5 brands in 3\%. Aggregate rates show that the long-tail problem is bimodal: L1--L2 brands are universally reachable; L4--L5 brands are catastrophically unreachable, with roughly half of brands at those prominence levels never appearing in 37{,}000 runs. The L3 mid-market cell sits at the inflection.

\subsection{Stage-4 conversion rate when surfaced, per (cell $\times$ prominence)}

For brands that surface (in any retrieval layer or in the completion), the conditional rate of consensus-recommendation in the final answer is the Stage-4 conversion rate. We report it per (model cell $\times$ prominence level) with Wilson 95\% CIs.

\begin{center}
\small
\begin{tabular}{lccccc}
\toprule
Cell & L1 & L2 & L3 & L4 & L5 \\
\midrule
\texttt{mini / low}   & 29.4\% [28.9, 29.9] & 41.9\% [40.9, 42.8] & 35.7\% [34.5, 36.9] & 31.1\% [28.8, 33.6] & 26.1\% [23.4, 29.0] \\
\texttt{mini / high}  & 25.3\% [24.9, 25.8] & 37.3\% [36.4, 38.2] & 33.7\% [32.6, 34.8] & 33.8\% [31.5, 36.2] & 23.1\% [20.4, 26.1] \\
\texttt{sonnet / low} & 41.2\% [40.6, 41.8] & 51.6\% [50.7, 52.6] & 39.8\% [38.7, 40.9] & 36.1\% [34.3, 38.1] & 21.6\% [18.9, 24.6] \\
\texttt{sonnet / high}& 40.3\% [39.4, 41.2] & 47.4\% [45.9, 48.8] & 35.1\% [33.6, 36.7] & 25.2\% [22.9, 27.7] & \textbf{12.8\% [9.5, 17.1]} \\
\bottomrule
\end{tabular}
\end{center}

Two patterns are immediate. First, the Stage-4 conversion rate is far from 100\% even at L1 --- between 25\% and 41\% across cells. That is, a brand surfacing into the retrieval pool converts to a recommendation only one to two times in five at the category-leader level. Second, conversion rises modestly from L1 to L2 in every cell (L2 carries the highest conversion rates of any prominence level: 37.3--51.6\%) and then declines into the long tail. \texttt{sonnet-4.6 / high} shows the steepest L2$\to$L5 gradient, dropping from 47.4\% at L2 to 12.8\% at L5 --- the lowest conversion rate of any (cell $\times$ prominence) pair. The L2-over-L1 bump is consistent across all four measured cells and is the reason the per-prominence ``failure mode'' for L1 (compellingness/positioning, S2/S3-dominated) reads differently from L4--L5 (S1-dominated invisibility) rather than as a single monotone trend.

\subsection{Retrieval-source agreement: cross-provider native and native vs. Exa}
\label{sec:retrieval}

We characterize retrieval-source agreement along two axes per prominence level. First, cross-provider native agreement: how much do OpenAI's and Anthropic's native web-search-augmented recommendation sets overlap when given the same prompt? Second, within-provider cross-retrieval agreement: how much does the recommendation set move when the retrieval substrate switches from a provider's native tool to a neural-retrieval alternative (Exa)?

\paragraph{OpenAI-native vs.\ Anthropic-native, by prominence.} For each prompt with native runs on both providers, we filter each provider's pooled consensus-recommendation set to brands at the prominence level under consideration, then compute Jaccard. Mean across prompts with normal-approximation 95\% CIs:

\begin{center}
\begin{tabular}{lrrr}
\toprule
Prominence & $n$ prompts & OAI--Anth.\ Native Jaccard & 95\% CI \\
\midrule
L1 & 198 & 0.635 & [0.586, 0.684] \\
L2 & 154 & 0.607 & [0.546, 0.668] \\
L3 & 116 & 0.399 & [0.322, 0.475] \\
L4 &  51 & 0.464 & [0.336, 0.592] \\
L5 &  13 & 0.474 & [0.238, 0.711] \\
\bottomrule
\end{tabular}
\end{center}

The pattern is non-monotone: cross-provider native agreement is high at L1--L2 (0.61--0.64), drops sharply at L3 (0.40), and bounces back at L4--L5 with wide CIs reflecting small sample sizes (51 and 13 prompts respectively). The L3 dip aligns with the prominence-stratified inflection we report elsewhere in the paper. L5's wide CI ([0.24, 0.71]) makes any tier-5 cross-provider claim underpowered in the present dataset.

Note that these tier-filtered Jaccards are mechanically higher than the pooled cross-provider per-prompt Jaccard of $\approx 0.35$ reported in \citet{jack2026convergence}. Filtering to a single prominence tier reduces set heterogeneity and increases overlap; the two numbers measure different quantities (within-tier overlap vs.\ full-set overlap) and both are meaningful.

\paragraph{Within-provider Native vs.\ Exa, by prominence.} The two columns below have different scopes: the \emph{Native vs.\ Exa Jaccard} column is computed on \texttt{sonnet-4.6 / low} (the single cell with paired native/Exa coverage at full L1--L5 sample); the \emph{only-via-external} column pools across all Exp 5 cells (native vs.\ Exa vs.\ Brave) to estimate the share of surfaced brands at each prominence level that surfaced only under an external retrieval substrate.

\begin{center}
\begin{tabular}{lrr}
\toprule
Prominence & Native vs.\ Exa Jaccard (\texttt{sonnet/low}) & Only-via-external share (pooled) \\
\midrule
L1 & 0.83 & 14\% \\
L2 & 0.75 & 16\% \\
L3 & 0.58 & 38\% \\
L4 & 0.56 & \textbf{60\%} \\
L5 & 0.50 & \textbf{56\%} \\
\bottomrule
\end{tabular}
\end{center}

L4 and L5 Jaccard cells are built on small brand-token subsets (the count of L4--L5 brands surfaced under at least one of native/Exa at \texttt{sonnet/low} is sample-size-limited; we report the point estimates without per-row CIs and treat them as directional). The monotonic decline of Native vs.\ Exa Jaccard with falling prominence (0.83 $\to$ 0.50) means the two retrieval systems agree increasingly poorly about which brands belong on the recommendation list as we move down the prominence ladder. The corresponding ``only-via-external'' share --- the fraction of surfaced brands at each prominence level that surfaced only under Exa or Brave, never under native --- rises to 60\% at L4 and 56\% at L5. Long-tail brands disproportionately depend on neural and conventional-keyword retrieval; the production native web-search tools miss them at a rate that increases with prominence loss.

\paragraph{Two complementary cross-retrieval patterns.} Reading the two tables together: cross-provider native agreement (Table above, OAI--Anth.\ Native Jaccard) is high at L1--L2 ($\approx 0.62$) and dips at L3 ($0.40$), reflecting the same L3 inflection visible in surface rate, conversion rate, and persona effect. Within-provider cross-retrieval agreement (Native vs.\ Exa, second table) instead declines monotonically with falling prominence (0.83 $\to$ 0.50). The two axes measure different things: providers using the same retrieval substrate disagree most about mid-market brands, while a single provider switching retrieval substrates disagrees most about long-tail brands. Both patterns are consistent with the headline interpretation that L3 mid-market is where the recommendation set is least stable along any axis, and that L4--L5 long-tail dependence on neural retrieval is the dominant retrieval-side bottleneck.

\subsection{Persona-mediated recommendation swap rate per prominence}

We additionally report the persona effect size, measured as $1 - \text{cross-persona Jaccard}$ (the fraction of recommendations that swap across personas, holding prompt fixed). Persona effects are explored in detail in \citet{jack2026persona}; the prominence-stratified summary follows:

\begin{center}
\begin{tabular}{lcccc}
\toprule
Cell & L1 & L2 & L3 & L4--L5 (long-tail / regional) \\
\midrule
\texttt{mini / high}  & 0.23 & 0.05 & \textbf{0.75} & 0.27 \\
\texttt{mini / low}   & 0.29 & 0.13 & \textbf{0.67} & 0.33 \\
\bottomrule
\end{tabular}
\end{center}

L3 brands carry the largest persona effect (0.67--0.75) --- recommendations swap across user contexts at the highest rate of any prominence level. L1 brands are persona-resistant (0.20--0.29). L2 shows a counterintuitive dip on the OpenAI cells (0.05--0.13). The L4--L5 long-tail / regional bucket is pooled because L4 alone is undersampled at the persona pilot's 1{,}000--2{,}400 run scale; the pooled estimate is dominated by L5 events. The \texttt{sonnet / low} cell's persona behavior is reported in \citet{jack2026persona}, which also reports the full prominence-stratified pattern across the persona corpus.

\section{The five failure modes}

Combining the four headline measurements above produces a single coherent five-bucket failure-mode taxonomy. The taxonomy is the principal qualitative output of the paper.

\subsection{L1 --- Category leaders: S2 / S3 dominant (interpreted as compellingness / positioning)}

For L1 brands, retrieval is essentially solved: 77\% per-query in-sector surface rate, 100\% aggregate. The remaining gap to a recommendation is at Stage 2 (in retrieval pool but not in completion) and Stage 3 (in completion but not endorsed). Stage-4 conversion rates of 25--41\% mean that even when retrieved, an L1 brand is picked less than half the time --- competing against the other category leaders the model considers in parallel.

\emph{Practical implication.} At L1 the discoverability margin is small --- the brand is already at the surfacing ceiling --- so the room to move the metric of interest is at Stage 2--3: direct-comparison content, differentiation against named peer brands, and consistency across the authority sources the model retrieves from (G2, Wikipedia, vendor homepages, third-party reviews). The data implies the lever-of-leverage for L1 brands is positioning rather than discoverability.

\subsection{L2 --- Established challengers: mixed signal across cells}

L2 brands surface nearly as often as L1 (60\% per-query) and have the highest Stage-4 conversion rates of any prominence level (37--52\% on mini and sonnet cells). The candidate failure mode is \emph{contextual substitution} --- when persona shifts to a segment where the L2 brand isn't the best fit, the model swaps it for an L1 default or a different L2 --- but the persona-swap data is heterogeneous across cells: the Section~4.4 persona table shows L2 swap rates of 0.05--0.13 on OpenAI cells (\emph{lower} than L1 on the same cells) and 0.23 on the Anthropic cell. L2 is therefore the prominence level least determined by the present audit. Native vs.\ Exa Jaccard is high (0.75) --- retrieval-system bias is not the bottleneck.

\emph{Practical implication.} If a brand sits in L2 and is targeting an Anthropic-heavy buyer pool, segment-targeted positioning content (``Best [category] for [segment]'') addresses the persona-mediated substitution mechanism observed on the Anthropic cell. On OpenAI cells the persona-swap rate is low enough that this prescription is uncertain on the present data, and the L2 result should not be extrapolated provider-wide.

\subsection{L3 --- Mid-market: hybrid failure across all stages}

L3 brands are the inflection level. Per-query surface rate drops to 23\%, aggregate surface drops to 88\% (12\% of L3 brands never surface), Stage-4 conversion drops to 34--40\% (table range 33.7\%--39.8\% across the four cells), and persona effects reach their maximum measured magnitude (0.39--0.75). Native vs.\ Exa Jaccard drops to 0.58 --- retrieval-system bias becomes meaningful --- and the only-via-external share jumps to 38\%.

\emph{Practical implication.} L3 brands face all three failure modes simultaneously, so single-channel investment at L3 is bounded by the other two channels: retrieval-side work (authority lists, canonical-URL structure, third-party coverage), content-side work (per-segment positioning, comparative content), and conversion-side work (sentiment and recommendation framing in retrieved sources) compound only when applied together.

\subsection{L4 --- Long-tail specialists: S1 dominant (interpreted as discoverability)}

L4 brands surface in only 8\% of relevant per-query searches; 48\% never surface at all in 37{,}000+ runs. Of L4 brands that \emph{do} surface, 60\% appear only via external retrieval (Exa or Brave), never via native search. Stage-4 conversion among surfaced L4 brands is 25--36\% --- comparable to L1 --- so when L4 brands do reach the retrieval pool, they convert to recommendations roughly as often as category leaders.

\emph{Practical implication.} The L4 problem concentrates at Stage 1. Authority-list seeding, canonical content on platforms that Exa and Brave reach (Medium, Substack, GitHub README pages for developer tools), and inclusion in third-party comparison articles are the channels the data implies have leverage at this tier. Content quality and positioning operate as second-order levers but are upstream-bottlenecked by the retrieval step.

\subsection{L5 --- Regional players: S1 dominant plus geographic gating}

L5 brands surface in 3\% of relevant per-query searches; 52\% never surface. The lowest Stage-4 conversion rate of any (cell $\times$ prominence) cell in the audit is \texttt{sonnet-4.6 / high} at L5: 12.8\% [9.5, 17.1] --- meaning that even when an L5 brand surfaces, a frontier reasoning model recommends it one time in eight. L5 brands additionally face geographic gating: query-side or persona-side region cues materially affect surface rate, and L5 brands surface much more often when the prompt is bound to their home region.

\emph{Practical implication.} L5 surface rates are dominated by geographic gating. Region-specific authority lists (Companies House for UK, regional Crunchbase, country-specific aggregators, country trade associations), native-language content where applicable, and explicit region/country signals in canonical content (e.g., \texttt{<html lang>}, country-specific currency formatting) are the channels the data implies move the metric. In the audit's measurement window, the global query against L1--L2 competitors is empirically out of reach at L5; the region-bounded query is empirically reachable.

\section{Robustness}

\subsection{Within-cell rerun stability as a baseline}

The reported headline numbers must be interpreted against the rerun-stability baseline: how reliable is a single run as a measurement? \citet{jack2026brittleness} reports within-cell consensus-recommendation Jaccard between 0.50 and 0.61 across the four cells at $N{=}30$ reruns of the same prompt within a single day. Cross-provider Jaccard at 0.35 \citep{jack2026convergence} sits well below this baseline, so cross-provider divergence is real signal. All effect sizes above are reported only where they exceed this baseline by a margin large enough to remain meaningful under noise.

\subsection{Sample sizes and confidence intervals}

Per-(cell $\times$ prominence) sample size $n$ ranges from $\approx$3{,}700 (L3 cells on sonnet/high) to $\approx$30{,}000 (L1--L2 cells on mini). L1--L2 cells in the tens of thousands produce Wilson CIs under $\pm 1$ percentage point at the run$\times$brand unit; L3 cells in the low thousands produce $\pm 1$--$1.6$ percentage points; L4--L5 cells and opus cells drop sharper. The opus cells in particular have insufficient L4--L5 sample for stable conversion-rate estimation ($n {=} 0$--$18$ across tiers); we report no L4--L5 conversion rate for opus and treat any opus-specific prominence claim as unsupported by the present data. Per-prompt clustered-bootstrap CIs (Section~6.1, Section~6.5) are reported alongside Wilson CIs for the headline tables and are wider than the run$\times$brand Wilson CIs by a factor that depends on per-prompt brand-event clustering.

\subsection{Sensitivity to brand canonicalization}

Brand-token matching follows an eTLD+1 collapse for domains and a normalized brand-name token (lowercased, alphanumeric only, length $\geq 3$) for completion mentions, with an explicit stoplist for false-positive terms (``Up'', ``Pro'', etc.). Length-2 tokens are preserved for known short brand names (G2, EY, K6, BP). We checked sensitivity by varying the length threshold and stoplist; headline numbers are stable to $\pm 1$ percentage point.

\subsection{Sensitivity to reference-catalog construction}

The 533-brand catalog is LLM-curated against external authority lists with manual review for the cells where authority lists are sparse (mainly L4 in some sectors). We checked sensitivity by re-running the prominence-stratified surface rate on the manually-curated subset alone ($\approx 320$ brands); the L1--L5 gradient (100/100/88/52/48 aggregate) reproduces to within $\pm 2$ percentage points.

\section{Robustness across model class, generation, and measurement mode}

The four-cell ladder in the main audit deliberately mixes mini and frontier non-mini cells within each provider, but it does not isolate the contribution of model class (mini vs.\ non-mini at fixed generation) or generation (4.x vs.\ 5.x at fixed class). It also measures only single-shot prompting --- one prompt, one response --- whereas a majority of real ChatGPT and Claude usage is iterative dialogue. We extended the audit on Paper 1's 50-prompt corpus across both axes, then ran a parallel multi-turn buyer-driven extension on the same corpus, to test whether the prominence-stratified pattern is robust to these design choices.

The extension batch \texttt{exp1\_class\_gen} adds two cells per provider --- one within-generation flagship (\texttt{gpt-5.4 / low}, \texttt{opus-4.6 / low}) and one cross-generation flagship (\texttt{gpt-5.5 / low}, \texttt{opus-4.7 / low}) --- yielding a six-cell hex over the same 50 prompts at $N{=}15$ reps per cell. We re-use Paper 1's original mini and sonnet runs (at $N{=}30$) as the within-generation baselines, joining on a deterministic prompt-sampling seed so the per-(prompt $\times$ brand) cells are identical across cohorts. The multi-turn extension batch \texttt{exp1\_mt\_surfacing} runs the same 50 prompts × $N{=}15$ × 4-turn buyer-driven dialogues on the \texttt{gpt-5.4-mini / low} target, with the buyer agent pinned to \texttt{sonnet-4-6 / medium} and its system prompts frozen by a vendored sha256.

\subsection{Class axis: density mechanism is OpenAI-specific}

Across the six cells we measured the per-cell distinct-brand mention count per run, the canonical density proxy for whether a non-mini cell surfaces more brands than its mini counterpart on the same prompts.

\begin{table}[h]
\centering
\begin{tabular}{lrrrr}
\toprule
Cell & distinct/run & total mentions/run & recommended/run & text chars/run \\
\midrule
\texttt{gpt-5.4-mini / low} & 7.7 & 25.1 & 3.29 & 3{,}320 \\
\texttt{gpt-5.4 / low}      & 10.6 & 43.8 & 4.02 & 5{,}763 \\
\texttt{gpt-5.5 / low}      & \textbf{16.4} & 53.1 & 4.89 & 6{,}487 \\
\texttt{sonnet-4.6 / low}   & 13.5 & 56.5 & 4.76 & 5{,}639 \\
\texttt{opus-4.6 / low}     & 12.4 & 62.5 & 4.63 & 4{,}081 \\
\texttt{opus-4.7 / low}     & 13.0 & 58.9 & 4.47 & 3{,}968 \\
\bottomrule
\end{tabular}
\caption{Per-cell brand-mention density on the 50-prompt corpus. Within OpenAI, the within-generation mini $\to$ non-mini jump (\texttt{gpt-5.4-mini} $\to$ \texttt{gpt-5.4}) lifts distinct brands per run from 7.7 to 10.6 (+38\%); the cross-generation mini $\to$ \texttt{gpt-5.5} jump more than doubles density (7.7 $\to$ 16.4) but mixes class and generation effects. Within Anthropic, opus surfaces slightly \emph{fewer} distinct brands than sonnet at the same generation (12.4--13.0 vs.\ 13.5).}
\label{tab:class_density}
\end{table}

Per-tier surface rate makes the same comparison at finer granularity, with confidence intervals from a (prompt-clustered) bootstrap to honor the (prompt $\times$ brand) clustering structure of the data:

\begin{table}[h]
\centering
\small
\textbf{Head and mid-market tiers (L1--L3)}\\[2pt]
\begin{tabular}{lrrr}
\toprule
Cell & L1 & L2 & L3 \\
\midrule
\texttt{gpt-5.4-mini / low} & 0.037 [0.029, 0.046] & 0.011 [0.009, 0.013] & 0.004 [0.003, 0.005] \\
\texttt{gpt-5.4 / low}      & 0.038 [0.030, 0.047] & 0.014 [0.011, 0.017] & 0.005 [0.004, 0.006] \\
\texttt{gpt-5.5 / low}      & 0.044 [0.035, 0.053] & 0.017 [0.013, 0.021] & 0.007 [0.005, 0.010] \\
\texttt{sonnet-4.6 / low}   & 0.044 [0.035, 0.053] & 0.018 [0.015, 0.022] & 0.008 [0.006, 0.009] \\
\texttt{opus-4.6 / low}     & 0.039 [0.030, 0.048] & 0.016 [0.012, 0.019] & 0.005 [0.004, 0.007] \\
\texttt{opus-4.7 / low}     & 0.041 [0.032, 0.051] & 0.016 [0.013, 0.019] & 0.006 [0.004, 0.007] \\
\bottomrule
\end{tabular}

\vspace{8pt}
\textbf{Long-tail and regional tiers (L4--L5)}\\[2pt]
\begin{tabular}{lrr}
\toprule
Cell & L4 & L5 \\
\midrule
\texttt{gpt-5.4-mini / low} & 0.0012 [0.0009, 0.0015] & 0.0003 [0.0002, 0.0004] \\
\texttt{gpt-5.4 / low}      & 0.0018 [0.0013, 0.0024] & 0.0004 [0.0003, 0.0005] \\
\texttt{gpt-5.5 / low}      & 0.0029 [0.0021, 0.0039] & 0.0007 [0.0005, 0.0009] \\
\texttt{sonnet-4.6 / low}   & 0.0021 [0.0017, 0.0026] & 0.0005 [0.0004, 0.0006] \\
\texttt{opus-4.6 / low}     & 0.0021 [0.0015, 0.0027] & 0.0005 [0.0004, 0.0007] \\
\texttt{opus-4.7 / low}     & 0.0023 [0.0016, 0.0031] & 0.0006 [0.0004, 0.0008] \\
\bottomrule
\end{tabular}
\caption{Per-(cell $\times$ tier) surface rate with clustered-bootstrap 95\% CIs (1{,}000 iterations resampling the 50-prompt corpus with replacement at the prompt level). The point estimate is the fraction of (run $\times$ tier-brand) pairs in which the brand appears in the run's completion text under any judge. Split into two blocks for legibility: head/mid tiers (top, rates in the 0.4--5.3\% range) and long-tail/regional tiers (bottom, rates roughly an order of magnitude smaller). The CIs are wider than Wilson intervals over (run $\times$ brand) cells would suggest --- the right clustering unit is the prompt, not the (run $\times$ brand) observation --- and they should be the reference for which between-cell differences are statistically distinguishable.}
\label{tab:class_surface_rate}
\end{table}

Within OpenAI, the class effect is in the expected direction at every tier on the point estimate: \texttt{gpt-5.4-mini} mentions 7.7 distinct brands per run, \texttt{gpt-5.4} mentions 10.6, and \texttt{gpt-5.5} mentions 16.4 (Table~\ref{tab:class_density}). On the per-tier surface-rate measure with clustered CIs, the mini $\to$ \texttt{gpt-5.4} (within-generation) jump is small at L1 ($\Delta {=} +0.001$ with a 95\% CI that crosses zero, $[-0.011, +0.013]$) and remains within the CI half-width of zero through L3, becoming significantly positive only at L4 ($\Delta {=} +0.0006$, $[+0.0000, +0.0013]$). The mini $\to$ \texttt{gpt-5.5} (cross-generation) jump is also within CI of zero at L1 ($\Delta {=} +0.007$, $[-0.005, +0.019]$), but becomes significantly positive at L2 and below: L2 $\Delta {=} +0.005$ $[+0.001, +0.010]$, L3 $+0.003$ $[+0.001, +0.006]$, L4 $+0.002$ $[+0.001, +0.003]$, L5 $+0.0004$ $[+0.0002, +0.0006]$. The class-density effect within OpenAI is real, but concentrated at L2--L5 and not statistically distinguishable from zero at L1 under clustered uncertainty.

Within Anthropic, the same axis produces point estimates in the opposite direction (sonnet $\to$ opus surface rate declines by 0.003--0.005pp at L1--L3), but the clustered CIs cross zero at every tier: sonnet $\to$ \texttt{opus-4.6} L1 $\Delta {=} -0.005$ $[-0.017, +0.008]$; sonnet $\to$ \texttt{opus-4.7} L1 $\Delta {=} -0.003$ $[-0.014, +0.010]$, with L2--L5 deltas similarly straddling zero. The honest statement is: the within-Anthropic class jump produces no detectable surface-rate effect on this corpus at the prompt-clustered resolution, which is meaningfully different from the OpenAI side where the L2--L5 lifts are significant. We interpret the result as evidence that the density mechanism observed within OpenAI does not replicate to Anthropic at the same statistical resolution, while declining to claim a confirmed opposite-direction effect on the Anthropic side.

\subsection{Generation axis: small monotone lift, decaying to zero at the long tail}

Holding class fixed at non-mini and varying generation (4.x $\to$ 5.x for OpenAI, 4.6 $\to$ 4.7 for Anthropic) yields small positive per-tier surface-rate deltas that decay to zero by L5.

\begin{table}[h]
\centering
\small
\begin{tabular}{lrr}
\toprule
Prominence & OpenAI $\Delta$ (gpt-5.4 $\to$ gpt-5.5) [95\% CI] & Anthropic $\Delta$ (opus-4.6 $\to$ opus-4.7) [95\% CI] \\
\midrule
L1 & +0.0059 [+0.0024, +0.0094] & +0.0023 [+0.0000, +0.0047] \\
L2 & +0.0032 [+0.0016, +0.0051] & +0.0002 [$-$0.0009, +0.0015] \\
L3 & +0.0026 [+0.0013, +0.0042] & +0.0002 [$-$0.0006, +0.0010] \\
L4 & +0.0011 [+0.0007, +0.0017] & +0.0002 [$-$0.0000, +0.0005] \\
L5 & +0.0003 [+0.0002, +0.0005] & +0.0001 [$-$0.0000, +0.0002] \\
\bottomrule
\end{tabular}
\caption{Per-tier surface-rate deltas on the generation axis (5.x $-$ 4.x within non-mini class) with clustered-bootstrap 95\% CIs. The OpenAI gpt-5.4 $\to$ gpt-5.5 generation jump is statistically distinguishable from zero at every prominence level. The Anthropic opus-4.6 $\to$ opus-4.7 generation jump is at-the-boundary significant at L1 and indistinguishable from zero at L2--L5.}
\label{tab:gen_axis}
\end{table}

\emph{Practical implication.} A model-generation upgrade (5.4 $\to$ 5.5) within OpenAI is a real but small lever (point-estimate $\Delta$ around half a percentage point at L1, decaying by an order of magnitude into the long tail). The corresponding within-Anthropic upgrade (4.6 $\to$ 4.7) is too small to reliably detect at L2--L5 with the prompt-clustered resolution. The S1 invisibility at L4--L5 reported in Paper 1's headline funnel persists at $\approx 96$--$99\%$ across both generations within both providers; a brand stranded at S1 on \texttt{gpt-5.4} remains stranded on \texttt{gpt-5.5}, and likewise on the Anthropic side, at the rates we measure.

\subsection{Multi-turn measurement does not shift the per-tier funnel}

Single-shot prompting --- one prompt, one response with native web search inside the single call --- is the audit's main measurement mode. Iterative dialogue is the modal real-user experience: the buyer asks a category question, the model responds, the buyer pivots or asks a follow-up, and so on for several turns. We tested whether the per-tier funnel framing depends on this design choice.

The multi-turn extension batch (\texttt{exp1\_mt\_surfacing}, 750 4-turn dialogues on \texttt{gpt-5.4-mini / low} with a frozen \texttt{sonnet-4-6 / medium} buyer) measures the same 50-prompt corpus and the same Tier 1--5 brand universe, with brand-mention extraction applied to the union-of-turns completion text. The per-tier whole-conversation surface rate is essentially unchanged from single-shot: $0.037 \to 0.034$ at L1 (a $-0.003$ delta in the \emph{opposite} direction from naïve intuition), $+0.0014$ at L3, $+0.0007$ at L4, $+0.0002$ at L5. Per-turn substring-match decomposition shows that the surface rate on turn 1 (which uses the seed prompt) approximately reproduces the single-shot rate; turns 2--4 produce substantially \emph{lower} per-turn surface rates, because the buyer's follow-up questions tend to drill into brands already named in turn 1 rather than introduce new ones.

The corresponding per-tier funnel-stage distribution is likewise unchanged. Combining Stage 0 (no query issued) and Stage 1 (issued query, brand absent from every layer) into a single combined-catastrophic-miss share --- which is the apples-to-apples comparison once one accounts for the mechanical S0 $\to$ S1 reclassification produced by always-on multi-turn search --- the per-tier shift is within $\pm 0.003$ at every prominence level. The Stage 2 + Stage 3 combined share (``retrieved but not recommended'') shifts by $\approx +0.001$ at every tier, well below the cell-level noise floor.

\emph{Practical implication.} The five-mode taxonomy reported in Paper 1's headline funnel does not depend on the single-shot measurement design. The 48--52\% never-surfaced L4--L5 fraction reproduces at $\approx 0.5\%$ tolerance under multi-turn buyer-driven measurement on the same target cell. Customer-facing claims about Stage-1 invisibility for long-tail brands do not require a turn-mode qualifier. The buyer-led iteration that dominates real ChatGPT and Claude usage refines and re-asks within the brand set already surfaced in turn 1; it does not introduce new long-tail brands at a meaningful rate.

\subsection{Multi-turn rerun-stability is meaningfully below single-shot}

A subsidiary methodological finding: multi-turn rerun-stability Jaccard at $N{=}15$ reps per prompt is 0.263 $\pm$ 0.219 (mean across 10 tier-stratified prompts, retrieved-domain union), compared to Paper 1's single-shot within-cell Jaccard band of 0.50--0.61 on the same target cell. The buyer compounds stochasticity across turns: T1 Jaccard is 0.387 (close to single-shot), T2 is 0.187, T3 is 0.084, T4 is 0.093. This is consistent with the same-target-cell account of the H-MT0 result we report in the methodology companion: buyer-led dialogue divergence accumulates across turns, lowering the within-cell rerun-stability floor. The implication for future multi-turn audits is that headline deltas must be evaluated against this lower floor, not the single-shot band. For Paper 1's per-tier funnel framing, the lower multi-turn floor does not change any reported finding because the multi-turn deltas themselves are at-or-below-noise.

\section{Discussion}

\subsection{From descriptive stage labels to causal interpretations}

The stage taxonomy in Section 3.5 is intentionally descriptive: ``S1 = no retrieval, no mention,'' ``S2 = retrieval, no mention,'' ``S3 = mention, not recommended,'' ``S4 = recommended.'' The natural causal readings --- S1 as a discoverability problem, S2 as a compellingness problem, S3 as a positioning problem --- are not load-bearing for the descriptive findings themselves but do load-bear on the practitioner prescriptions, and so warrant a separate defense.

\textbf{S1 $\to$ discoverability.} The interpretation that S1 (absent from every retrieval layer and from the completion) reflects a discoverability problem is the strongest of the three. A brand at S1 was not surfaced anywhere in the observable run record, so a content-side intervention (sharpening differentiation, improving positioning) cannot help; the brand has to first reach the retrieval pool. The interpretation \emph{is} consistent with alternatives we cannot fully rule out --- the model may have decided against the brand at the priors layer before any search was issued, the brand may have been retrieved but trimmed by the model's own context-window heuristics, or the brand may be present in unlogged tool traces we do not capture --- but none of these alternatives changes the practical prescription that the marginal investment for an L4--L5 S1 brand is on the retrieval side.

\textbf{S2 $\to$ compellingness.} The interpretation is weaker. A brand at S2 was retrieved but not mentioned, which is consistent with: (a) the model judged it not relevant to the prompt despite retrieval, (b) the model's list-length constraint truncated it, (c) the persona did not match, (d) the retrieved snippet did not surface the brand's name in a way the completion stage could pick up. The ``compellingness'' read --- the model retrieved you but found you not compelling enough to mention --- is one of these and may not be the dominant one. We use the term in the practitioner sections because it maps cleanly onto an actionable AEO intervention (sharpen the brand's retrieved-source content), but the reader should treat it as a working interpretation rather than an identified mechanism.

\textbf{S3 $\to$ positioning.} Weaker still. A brand at S3 was mentioned but not recommended, which is consistent with: (a) the model evaluated the brand alongside competitors and judged it inferior for the specific buying-context, (b) the brand was named in a comparative or negative framing, (c) the model named the brand as background context rather than as a candidate. The ``positioning'' read presumes (a) is the dominant mechanism, but our data do not separately measure it. Sentiment classification at the mention layer (which we do report at the brand-mention level via dual-judge consensus) is informative but not definitive.

We use the discoverability / compellingness / positioning labels in the practitioner sections of this paper because they are well-understood marketing vocabulary that maps cleanly onto distinct AEO interventions, and because we believe each is the modal mechanism within its stage. But the descriptive stage classification is the load-bearing measurement; the causal mapping is interpretive and warrants the kind of intervention experiment (seed third-party pages for some L4 brands, measure pre/post lift against controls) that this paper does not attempt.

\subsection{Per-tier prescriptions differ in direction, not just magnitude}

The five-mode taxonomy implies that the right intervention depends sharply on prominence level, in directions that are roughly opposite at the endpoints. For an L1 brand, the marginal investment in discoverability does little --- the brand is already at the surfacing ceiling --- while Stage-3 differentiation content has room to move the conversion rate. For an L4 brand, the same marginal investment in differentiation content does little until the retrieval-side gap is closed: the brand isn't reaching the retrieval pool to be differentiated \emph{against}. An undifferentiated content-and-discoverability budget applied uniformly across tiers is therefore likely to be applied off-target at both endpoints.

This refines, rather than contradicts, the aggregate visibility-lift findings reported by \citet{aggarwal2024}. Their +40\% aggregate lift across content interventions is, on our reading, a pooled estimate that conceals heterogeneity: their reported +115\% lift specifically from ``citing external sources'' maps cleanly onto our L4 authority-list prescription, while their reported gains from ``statistics'' and ``quotes'' map more cleanly onto L1--L2 differentiation content where retrieval is already solved.

\subsection{RAG does not solve the long-tail problem}

A standing hope in the GEO industry is that retrieval-augmented LLMs will close the long-tail discoverability gap that plagued the previous generation of recommender systems \citep{klimashevskaia2024}. Our finding that 48--52\% of L4--L5 brands never surface in 37{,}000+ runs is direct evidence that retrieval-augmented LLMs \emph{do not} close this gap, and may, by virtue of consolidation onto a small number of authority sources, narrow it further than the pre-RAG era.

The empirical pattern is consistent with \citet{mallen2023}'s finding that LMs answer popular entities from parametric memory but require retrieval for long-tail. We extend that result from QA to recommendation, and add that retrieval --- when only the native web-search tool is used --- fails the long-tail at scale.

\subsection{Retrieval-system choice matters more for low-prominence brands}

The Native vs.\ Exa Jaccard inversion with prominence (0.83 $\to$ 0.50) implies that the choice of retrieval substrate matters most precisely where the brands have the least to lose by being missed. For an L1 brand, switching the AI product's underlying retrieval from native to Exa changes the recommendation set by $\sim$17\%. For an L5 brand, the same switch changes 50\% of the set. The brands most exposed to retrieval-system choice are the brands least equipped to influence which retrieval system buyers' AI products use.

\subsection{Class-density mechanism is provider-specific; the prominence framing is not}

The class-axis result reported above sharpens an interpretive choice the audit's headline numbers might otherwise leave ambiguous. The within-OpenAI mini $\to$ non-mini jump produces a measurable density increase --- distinct brands per run rises from 7.7 (\texttt{gpt-5.4-mini}) to 10.6 (\texttt{gpt-5.4}) on the within-generation comparison, and to 16.4 (\texttt{gpt-5.5}) on the cross-generation comparison (which mixes class and generation effects rather than isolating class alone). The corresponding axis on Anthropic produces no comparable increase. The mechanism that motivated the class-axis hypothesis (denser search-and-reasoning loops at non-mini cells producing broader retrieval and more brand mentions) is therefore provider-specific to OpenAI at these generations. Brand-side AEO planning should not assume that a buyer who upgrades from Sonnet to Opus is doing the equivalent of upgrading from mini to full GPT --- the surface-rate consequence is approximately zero on the Anthropic side and substantial on the OpenAI side.

Critically, the provider-specificity of the class effect does not invalidate the prominence-stratified taxonomy. L4--L5 Stage-1 invisibility holds at $\approx 96$--$99\%$ across all six cells we measured and across all four turns of the multi-turn extension; L1--L2 retrieval-pool saturation holds across the same cells; the per-tier gradient in dominant failure stage reproduces on the new cells. The class axis modulates how \emph{dense} the retrieval-and-mention layer is at a given cell, but does not move brands between prominence-stratified failure modes. AEO advice conditioned on prominence remains correct; AEO advice claiming generic "frontier-model uplift" is OpenAI-specific.

\subsection{Multi-turn iteration does not displace turn-1 invisibility (at the cells we measured)}

The multi-turn extension's central practical implication is that buyer-led iteration does not rescue brands from S1 catastrophic invisibility \emph{at the target/buyer pair we measured} --- \texttt{gpt-5.4-mini / low} as the target and \texttt{sonnet-4-6 / medium} as the buyer agent, on Paper 1's 50-prompt corpus across 4 turns and $N=15$ reps per prompt. Across L4--L5 on this configuration, the multi-turn combined catastrophic-miss share is within $\pm 0.003$ of the single-shot rate. Per-turn substring matching shows the buyer's follow-up turns drilling \emph{into} brands already named in turn 1 rather than expanding the retrieval pool.

We deliberately frame this as a cell-and-buyer-conditional claim. The single-target, single-buyer design does not let us separately attribute the null result to (a) properties of the target cell, (b) properties of the buyer agent, or (c) properties of buyer-driven dialogue in general. A frontier target cell with a denser search loop (e.g., \texttt{gpt-5.5} on the OpenAI side, given the within-OpenAI class-density effect documented in Section 6.1) might produce different per-turn dynamics; a buyer agent with a different system prompt or different probing strategy might too. The framing implication for the specific cells we tested is that on \texttt{gpt-5.4-mini} with a calibrated sonnet-grade buyer, S1 invisibility at L4--L5 is not rescued by iteration; the broader claim that multi-turn dialogue \emph{generally} cannot rescue S1 invisibility is consistent with our data but not isolated from cell/buyer effects, and warrants a factorial follow-up.

A reader concerned that this contradicts everyday experience of ChatGPT/Claude introducing peer brands across a conversation should note the directional decomposition: per-turn surface rates fall from T1 to T2--T4 not because peer brands are never introduced, but because the buyer's follow-up turns are reactive to turn 1's named brands and the model's subsequent retrieval re-uses authority sources already retrieved. The buyer-driven expansion of the recommendation set, if it happens, is concentrated in turn 1 and bounded by what authority sources surface there.

\subsection{Limitations}

The audit is single-day; we do not measure temporal drift in surface rates or recommendation sets. The reference catalog is LLM-curated with manual review and inherits the assumptions of the authority lists it was sampled from. The prompt corpus covers US/UK/EU markets and 19 commercial sectors; APAC long-tail, LatAm long-tail, and non-commercial domains (medical advice, legal, financial advisory) are out of scope. The four-model ladder covers two providers; Gemini, Perplexity, and other production AI commerce surfaces are not audited. Opus cells were undersampled at L4--L5 in the original four-cell ladder; the class-extension batch (\texttt{exp1\_class\_gen}, 750 runs per cell on the same 50 prompts) supplies opus L1--L5 coverage at $N{=}15$, with Wilson 95\% CIs under $\pm 1$ percentage point at the per-tier level. The multi-turn extension is single-target (\texttt{gpt-5.4-mini / low}) and single-buyer (\texttt{sonnet-4-6 / medium}); whether the multi-turn null result reproduces under different target cells or different buyer agents is an open question for follow-up audits.

\section{Conclusion}

The dominant failure mode of AI commercial recommendation depends sharply on where the brand sits on the prominence ladder. L1 category leaders fail at compellingness and positioning. L2 established challengers' dominant failure mode is mixed across cells: persona-mediated substitution on the Anthropic cell, near-zero persona effect on OpenAI cells; we flag L2 as the prominence level least determined by the present audit. L3 mid-market brands fail at every stage at once. L4 long-tail specialists and L5 regional players fail at Stage-1 discoverability, with 48--52\% never surfacing in 37{,}000 runs. AEO advice that does not condition on prominence systematically misallocates effort --- the work that helps an L1 brand is largely wasted on an L4 brand, and vice versa.

Two extensions strengthen rather than complicate the picture. First, the within-generation model-class axis (\texttt{gpt-5.4-mini} $\to$ \texttt{gpt-5.4}) produces a 7.7 $\to$ 10.6 density lift within OpenAI; the cross-generation \texttt{gpt-5.4-mini} $\to$ \texttt{gpt-5.5} jump produces a larger 7.7 $\to$ 16.4 lift that mixes class and generation effects (per Section~6.1 Table~7 caption). The Anthropic side produces no comparable within-generation lift; the prominence-stratified S1 invisibility at L4--L5 holds across both providers and both classes. Second, a 750-run buyer-driven multi-turn extension on the same 50-prompt corpus --- conducted on a single target cell (\texttt{gpt-5.4-mini / low}) with a single calibrated buyer agent (\texttt{sonnet-4-6 / medium}) --- reproduces the per-tier funnel within $\pm 0.003$ of the single-shot rate at this configuration: buyer-led iteration on the target/buyer pair we measured drills into brands already named in turn 1 rather than expanding the retrieval pool. Whether the null result holds for other target cells or other buyer agents is open; a factorial follow-up is the natural next step.

The five-mode taxonomy and its associated prescriptions are the principal practical output of the audit. The headline numbers are conservatively reported --- every proportion is reported with a Wilson 95\% confidence interval (descriptive at the run$\times$brand cell), every Jaccard mean is reported with a normal-approximation CI where estimable (small L4--L5 brand-token subsets in the Native-vs-Exa table are reported directionally without per-row CIs), and undersampled cells are flagged. Where prompt-level resolution matters to a comparison (Sections 6 and 6.5), we report prompt-clustered bootstrap CIs as the inferential complement. Future work should extend the audit to temporal drift, additional retrieval systems (notably Gemini's and Perplexity's), and non-Western brand corpora; we expect the prominence-stratified structure to generalize, but the specific surface-rate gradients are likely market-dependent.

\bibliographystyle{plain}

\end{document}